\documentclass[aps,reprint,groupedaddress,superscriptaddress]{revtex4-1}
\usepackage{amsmath}
\usepackage{amssymb}
\usepackage{amsfonts}
\usepackage[dvipdfmx]{graphicx,color}
\usepackage{textgreek}
\usepackage{makecell,tabularx}
\setcellgapes{3pt}

\usepackage{amsfonts}
\usepackage[colorinlistoftodos]{todonotes}
\usepackage{multirow}

\usepackage[normalem]{ulem}
\usepackage{color}
\newcommand{\f}{\bf \textcolor[rgb]{0,0.4,0}}

\newcommand{\fp}{\bf \textcolor[rgb]{0.7,0,0}}

\begin{document}

\title{The triple Leidenfrost effect: preventing drops coalescence on a hot plate}

\author{F. Pacheco-V\'azquez}
\affiliation{Instituto de F\'isica, Benem\'erita Universidad Aut\'onoma de Puebla, A. P. J-48, Puebla 72570, Mexico \\
}

\author{J. L. Palacio-Rangel}
\affiliation{Instituto de F\'isica, Benem\'erita Universidad Aut\'onoma de Puebla, A. P. J-48, Puebla 72570, Mexico \\
}

\author{R. Ledesma-Alonso}
\affiliation{Universidad de las Am\'ericas Puebla, San Andr\'es Cholula, C.P. 72810, Puebla, Mexico. 
}

\author{F. Moreau}
\affiliation{Institut Pprime, UPR 3346 CNRS, ENSMA, Universit\'e de Poitiers, Futuroscope Cedex, France
}

\date{\today}

\begin{abstract}

We report on the collision-coalescence dynamics of drops in Leidenfrost state using liquids with different physicochemical properties. Drops of the same liquid deposited on a hot concave surface coalesce practically at contact, but when drops of different liquids collide, they can bounce several times before finally coalescing when the one that evaporates faster reaches a critical size, of the order of the capillary length. The bouncing dynamics is produced because the drops are not only in Leidenfrost state with the substrate, they also experience Leidenfrost effect between them at the moment of collision. This happens due to their different boiling temperatures, and therefore, the hotter drop works as a hot surface for the drop with lower boiling point, producing three contact zones of Leidenfrost state simultaneously. We called this scenario \textit{the triple Leidenfrost effect}. 

\end{abstract}


\maketitle

A liquid droplet deposited on a solid substrate considerably hotter than the liquid boiling temperature, $T_B$, levitates on its own vapor, reducing its evaporation rate and the friction with the substrate. This widely studied phenomenon is called Leidenfrost effect~\cite{leidenfrost1756aquae,gottfried1966leidenfrost,biance2003leidenfrost}. The minimum temperature $T_L$ required to observe such scenario depends on the liquid and substrate properties \cite{Bernardin2002,Kim2011,Nagai1996,Vakarelski2012}. For instance, a water drop enters in Leidenfrost state on a polished aluminum plate at $T_L\approx160-200^{\circ}$C \cite{Bernardin2002,hidalgo2016leidenfrost}. In contrast, an ethanol droplet exhibits the Leidenfrost transition on a heated oil pool when the difference in temperatures, $T_L - T_B$, is only of a few degrees~\cite{maquet2016leidenfrost}, revealing that  a slight superheating is enough to observe droplet levitation on a liquid substrate.

Recently, different dynamics involving the Leidenfrost effect have been explored: self-propelled droplets~\citep{lagubeau2011leidenfrost,Graeber2021}, sustained rotation~\citep{Agrawal2019,Bouillant2018}, star-shaped oscillations~\citep{burton2012geometry,hidalgo2016leidenfrost,Yi2020}, inverted Leidenfrost~\cite{Marston2012}, bouncing hydrogel balls \cite{waitukaitis2017coupling}, exploding droplets~\citep{Moreau2019}, etc. These studies suggest possible applications of the Leidenfrost mechanism in engineering and microfluidics. However, this can be challenging when droplets are manipulated simultaneously considering coalescence. Therefore, it is crucial to understand how Leidenfrost drops of different liquids coalesce.
When two droplets of the same liquid collide, they coalesce if the gas film between them is drained during the collision time. At room temperature, different regimes of bouncing, coalescence and separation in two-droplets collisions are known depending on the Weber number (the ratio of collision energy to surface energy), impact parameter (the deviation of  the droplets trajectories from that of head-on impact), and droplets size ratio~\cite{Qian1997,Tang2012,Krishnan2015,Diwari2020}. The dynamics also depends on liquids miscibility~\cite{Gao2005,Chen2006,Zhang2020}:  a water droplet  coalesces with an ethanol droplet~\cite{Gao2005} but it can bounce against an oil droplet~\cite{Chen2006}. On a superamphiphobic surface, two micrometric oil droplets may coalesce, bounce or propel, due to the transfer of surface energy to mechanical energy \cite{Olinka2020}. This energy transfer is also responsible of self-propelled jumping upon coalescence of water droplets in Leidenfrost state \cite{Liu2014}.

In this letter, we explore the collision-coalescence dynamics of two Leidenfrost drops of different liquids. First, we determined the Leidenfrost temperature $T_L$ for each liquid. Then, we focused on the collision of two large drops on a hot concave surface at temperature $T_s > T_L$.  When two drops collide, the outcome (direct coalescense or bouncing) depends on the difference in boiling temperatures, see Movie \cite{SI}. The bouncing dynamics can be explained assuming that the Leidenfrost effect also occurs between drops with large difference in boiling points, generating a vapor layer that prevents coalescence. After one droplet evaporates and reaches a size similar to its capillary length, the drops coalesce regardless of their initial volumes, impact velocity and substrate temperature. Finally, striking coalescence scenarios are observed.

\begin{figure*}
\includegraphics[width=17.5cm]{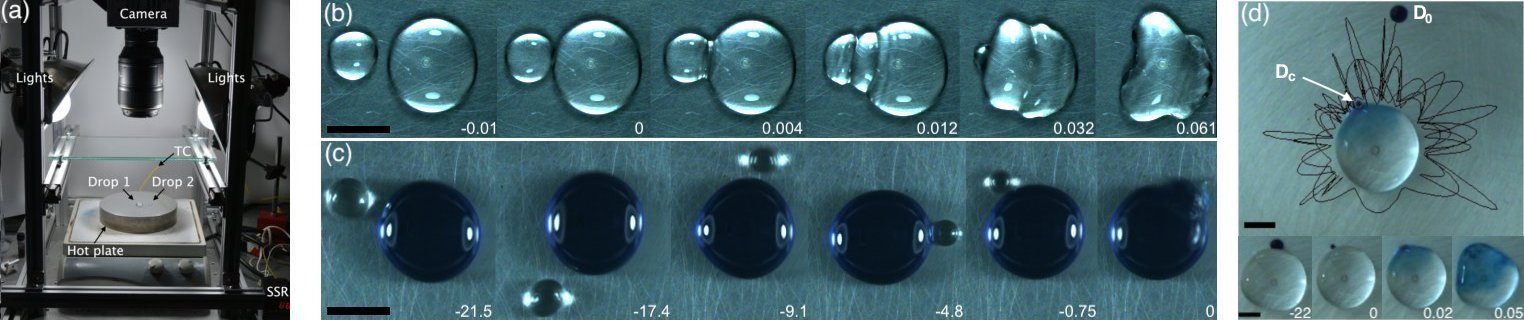}
\caption{a) Experimental setup. b) Direct coalescence of two water drops in Leidenfrost state. c) Consecutive bouncing of an ethanol droplet (transparent one) against a water droplet (tinted with methylene blue) during several seconds before coalescing. d) Path of an acetonitrile droplet (tinted in blue) bouncing several times against a water drop.  The snapshots show that the water droplet, initially transparent, turns bluish suddenly when the droplets coalesce. In (b-d), the elapsed time is indicated in seconds in each snapshot, with t=0 s corresponding to the moment of coalescence [scale bars: (b-c) 10 mm, (d) 5 mm].
}
\label{fig1}
\end{figure*}

\paragraph*{Experimental procedure.-} We placed on a hotplate a polished cylindrical aluminum substrate (15 cm diameter, 4 cm thickness, roughness $< 0.5~\mu$m) machined radially with a small angle ($\theta=2^\circ$), in order to keep the droplets at the center of the substrate (to the eye, the surface looks flat). The substrate temperature $T_s$ was controlled with a solid state relay (SSR) and monitored with a thermocouple (TC). Experiments were performed with one or two drops of initial volume $V_0\in[100,1000]~\mu l$ of 11 low viscosity liquids with different  physicochemical properties~\cite{SI}. The drops were deposited on the plate using micropipettes, and the dynamics was filmed from the top with a high speed camera (Fig. \ref{fig1}).

\begin{figure}[ht!]
\includegraphics[width=8.6cm]{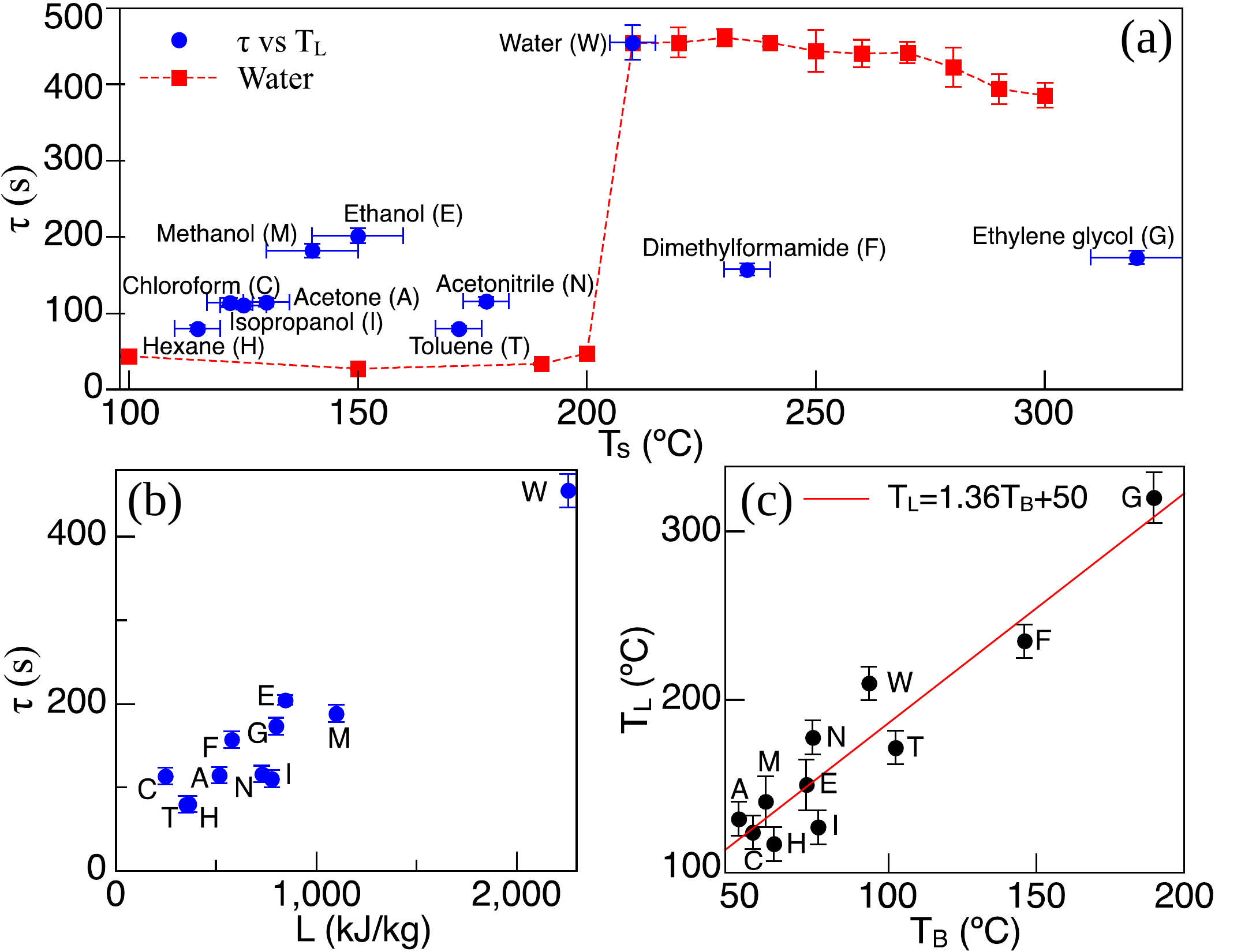}
\caption{a) Evaporation time $\tau~vs$ substrate temperature $T_s$ [red squares] and $\tau(T_L)$ [blue dots], for 0.5 ml drops of different liquids. b) $\tau~vs~L$.  c) $T_L~vs~T_B$ for the liquids in (a).  }
\label{fig2}
\end{figure}

\paragraph*{One droplet evaporation.-}Initial experiments were focused on determining the Leidenfrost temperature for each liquid by measuring the evaporation time $\tau$ of $500~\mu l$ droplets on the substrate at different temperatures $T_s$. An abrupt increase of $\tau(T_s)$ occurs at $T_s \approx T_L$ \cite{biance2003leidenfrost}. In Fig. \ref{fig2}a,  the dependence of $\tau$ on $T_s$ is reported for water (red squares), in which case $T_L \approx 210^{\circ}$C.  Similar plots were obtained for all liquids.  For clarity, only the values of $\tau(T_L)$ are shown in Fig. \ref{fig2}a (blue points). In most cases $\tau(T_L)\sim 100-200~\rm{s}$, but for water $\tau(T_L)\sim450~\rm{s}$. This larger evaporation time for water is related to its latent heat $L$, which is more than twice larger than the $L$ values of the other liquids (Fig. \ref{fig2}b). The Leidenfrost temperature found for each liquid was plotted as a function of the corresponding boiling temperature in Fig. \ref{fig2}c. Although finding a precise dependence of $T_L$ vs $T_B$ is not straightforward, the linear trend shown (red line) may provide a rough estimate of $T_L$ for an additional liquid deposited on the substrate, if its boiling temperature is known.

\paragraph*{Collision of two drops.-}Once known $T_L$ for each liquid,  two drops of different liquids (labeled Drop1 and Drop2 in Fig. \ref{fig1}a) with boiling temperatures $T_{B1}$ and $T_{B2}$ were deposited on the aluminum plate at $T_s=250^{\circ}$C ($T_s> T_L$ for both drops \cite{Note1}). First, we placed Drop1 of volume $V_0=1000~\mu l$ at the center of the plate. Then, Drop2 of $V_0\in[100,700]\mu l$ was released close to the edge of the plate at $d\sim$7 cm from Drop1. Drop2 slides under the action of gravity $g$ until colliding with Drop1 with impact velocity $v \sim \sqrt{2g d \sin \theta} \approx 22~$cm/s. The collision is followed by two dynamics depending on the liquids: i) direct coalescence, or ii) consecutive rebounds before coalescence. The direct coalescence lasts some milliseconds [Fig. \ref{fig1}b], and it was observed mainly with drops of the same liquid (e.g. water-water) or liquids with similar properties (e.g. ethanol-isopropanol). In contrast, drops with large  differences in properties (e.g.  water-ethanol or water-acetonitrile) remain bouncing during several seconds, or even minutes, while they evaporate until reaching a critical size to finally coalesce [Figs. \ref{fig1}c-d]. 

\begin{table}[h]
\includegraphics[width=8.7cm]{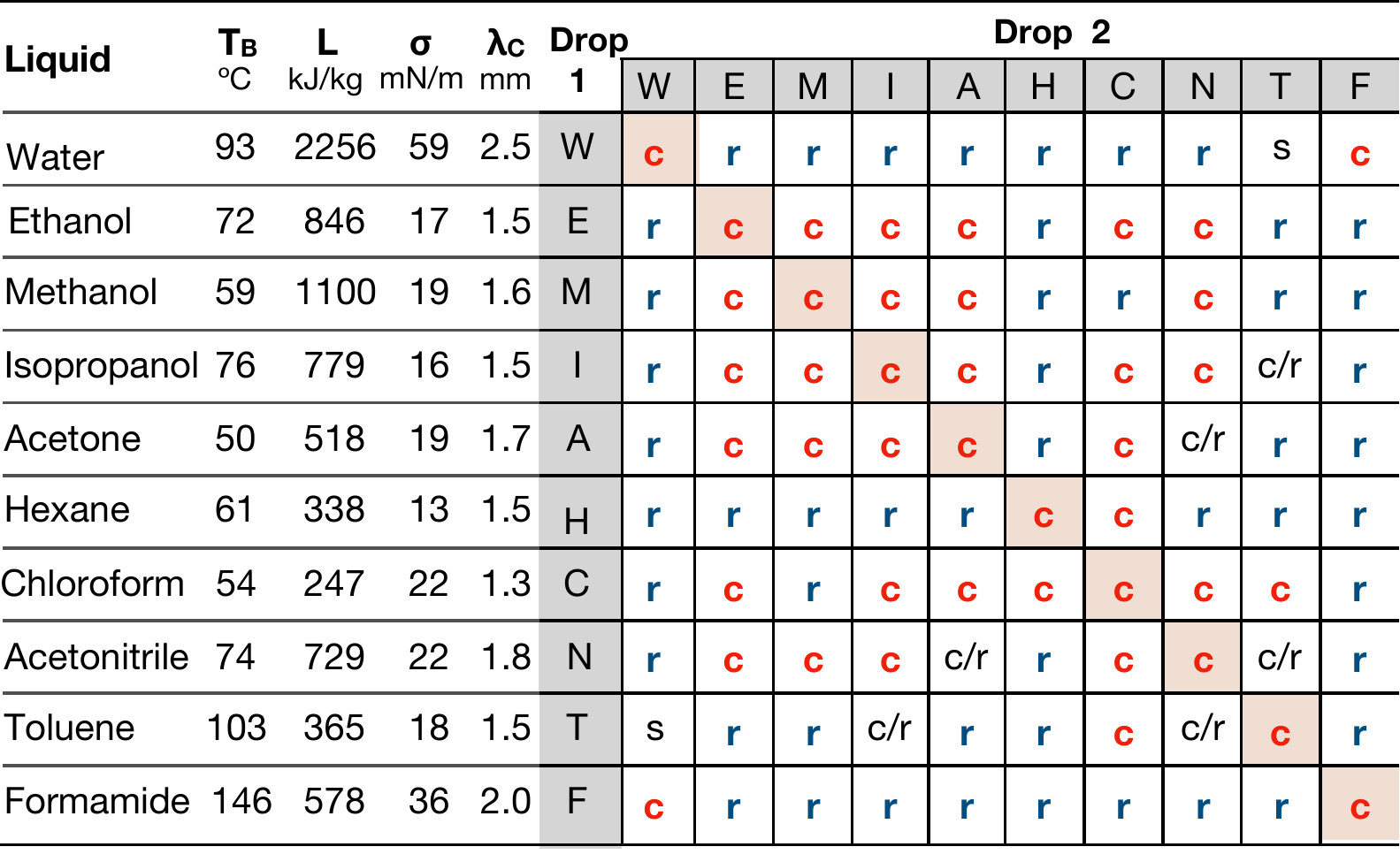}
\caption{\small Outcome of the collision of two Leidenfrost drops depending on the liquids pairs \cite{Note2}:  direct coalescence ({\fp{c}}), consecutive rebound ({\f{r}}), both scenarios (c/r), or separated phases (s). The latent heat $L$, surface tension $\sigma$ and capillary length $\lambda_c$ of each liquid at the boiling temperature $T_B$ are indicated (values in Puebla, Mexico, at elevation $\sim2200$ m above the sea level).  Data based on Refs. \cite{Vazquez1995,Yaws1999,Engtoolbox2001,Merck2020}.}
\label{Table1}
\end{table}
Table \ref{Table1} summarizes our observations for different liquids pairs considering at least five repetitions. Note that droplets of the same liquid coalesce directly (main diagonal), almost all liquids exhibit consecutive bouncing against water or dimethylformamide, and alcohols exhibit direct coalescence between them. Three pairs (acetonitrile with acetone or toluene, and toluene with isopropanol) can either coalesce or bounce a few times before coalescing, and toluene with water merge but remain as separated phases because they are immiscible.

\begin{figure}[ht!]
\includegraphics[width=8.6cm]{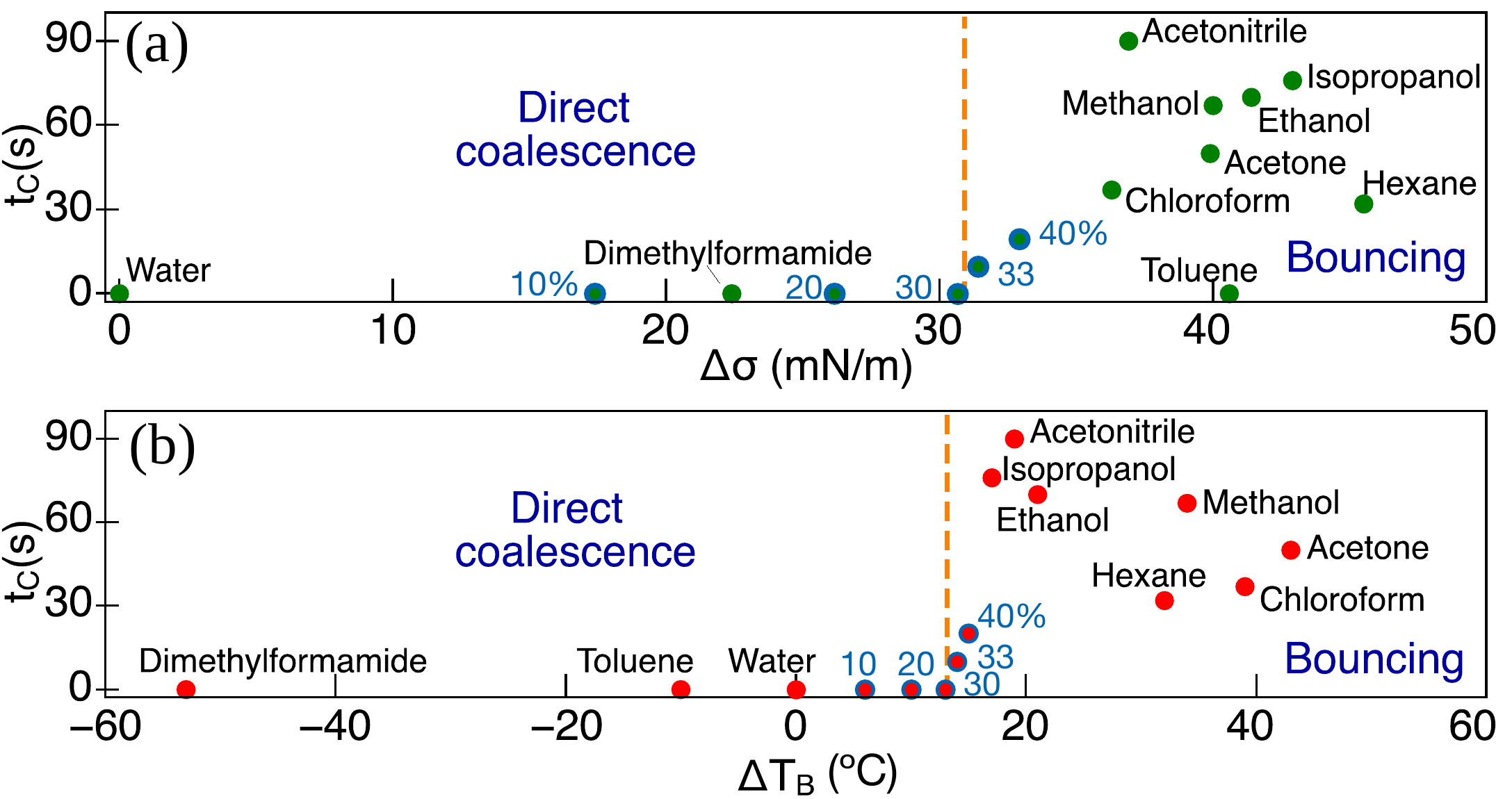}
\caption{\small Coalescence time $t_C$ for a Drop2 of $250~\mu$l of the indicated liquids with a 1 ml water Drop1 as a function of: a) $\Delta\sigma$, and b) $\Delta T$. Blue numbers denote the amount of ethanol ($\%$) in Drop2 of water-ethanol mixtures. }
\label{fig3}
\end{figure}

Since working with all the above combinations is complex due to the wide variety of liquid properties, we focus the following discussion in experiments with a Drop1 of $1~ml$ of water, interacting with a Drop2 of $250~\mu l $ of another liquid (their properties are labelled with a subindex $_1$ and $_2$, respectively). For these combinations, we measured the time $t_c$ from Drop2 deposition until its coalescence with Drop1. Then, we realized that two parameters could be possibly used to determine the conditions for direct coalescence $(t_c~\sim0)$ or bouncing $(t_c\gg 0)$: the difference of surface tensions $\Delta\sigma=\sigma_1-\sigma_2$, see Fig. \ref{fig3}a, or the difference in boiling temperatures $\Delta T=T_{B1}-T_{B2}$, see Fig. \ref{fig3}b. Note in these plots the transition from direct coalescence to bouncing at $\Delta\sigma\approx 31$ mN/m, and at $\Delta T\approx 15^{\circ}$C, respectively (orange lines).  Dimethylformamide coalesces directly with water ($t_{c}\approx0$) while acetonitrile remains bouncing more than 90 s. Additionally, we performed experiments using mixtures of water and ethanol for Drop2, reducing
progressively $\sigma_2\approx$ 59 mN/m  and $T_{B2}\approx 93^{\circ}$C for pure water, in which case Drop2 coalesces directly with Drop1 (same liquid), until attaining an ethanol concentration of $33\%$, at which Drop2 exhibits bouncing before coalescence. For this concentration, $\sigma_2\approx$ 28 mN/m  and $T_{B2}\approx 78^{\circ}$C  \cite{Vazquez1995}, which corresponds to  $\Delta \sigma \approx 31$ mN/m and $\Delta T \approx 15^{\circ}$C. These transition values are in agreement with those found using different liquids. One can notice in Fig. \ref{fig3} that toluene (with $t_c=0$ s) is beyond the transition for $\Delta \sigma$ but it is in the right side considering $\Delta T_B$. This suggests that $\Delta T_B$ determines if two droplets bounce or coalesce; but let us analyze the bouncing dynamics to get more insights about the dominant mechanism. 

\begin{figure*}[ht!]
\includegraphics[width=18cm]{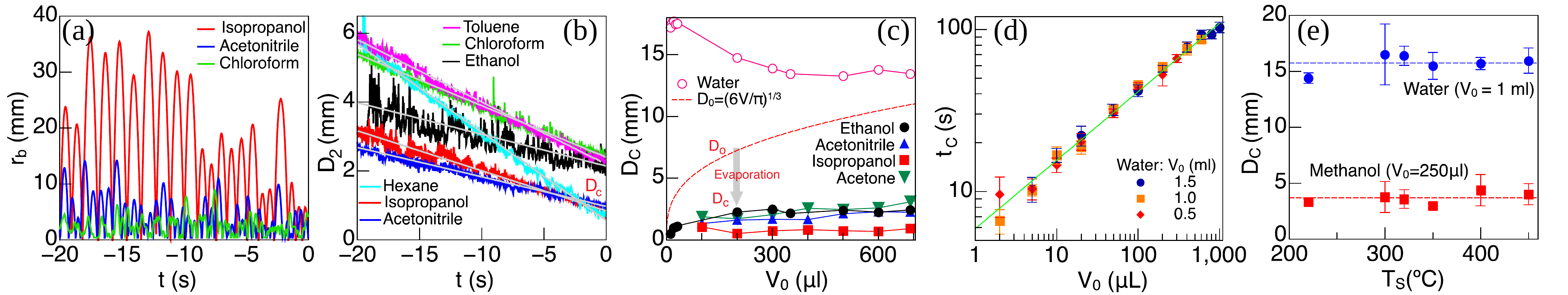}
\caption{ a) Size of radial bounces $r_b$ and b) diameter $D$ of Drop2 for different liquid drops during the last 20 s of bouncing before coalescence with water [the droplets coalesce at t=0 s, and only some liquids are shown for clarity].  c) Coalescence size $D_c$ of the satellite droplets as a function of the initial drop volume $V_0$.  The diameter of the central water droplet at the moment of coalescence is also indicated. d) Coalescence time $t_c$ of methanol droplets of different volumes $V_0$ with water droplets of three different initial volumes. e) Coalescence size $D_c$ of methanol and water drops as a function of the plate temperature $T_s$.}
\label{fig4}
\end{figure*}

\paragraph*{Bouncing dynamics and coalescence size:}
We performed particle tracking from videos taken at 60 fps focused on Drop2 (of different liquids) bouncing against Drop1 (water).  An example of this dynamics is shown in Fig. \ref{fig1}d for acetonitrile bouncing during 22 s before coalescing with water. The moment of coalescence corresponds to $t=0~\rm{s}$. For different liquids, Fig. \ref{fig4}a shows the radial position of Drop2, $r_b$, measured from the surface of Drop1: isopropanol performs the largest rebounds and chloroform the smallest ones. Moreover, the bouncing lengths are erratic, indicating that the rebound velocity can be higher than the impacting velocity. This is possible because the vapor layer is constantly replenished by the evaporation of the droplets \cite{Liu2014}. This also indicates that the moment of coalescence is independent of the impact velocity in the studied range. It is important to remark the clear contrast between our results with two Leidenfrost drops and those reported in the literature with drops at ambient temperature. Since the collisions are mainly radial in our experiments (impact parameter $X\sim 0$) and Drop2 bounces different radial distances (with $v\approx0-22$ cm/s), if one takes, for example, a Drop2 of ethanol (density $\rho=0.748$ g/cm$^3$) and diameter $D\sim 1.5$ cm,  the Weber number, We$= \rho v^2 D /\sigma$, is in the range $0<$We$<46$. At room temperature and similar Weber numbers, the collision of a water drop with an ethanol drop always results in coalescence  \citep{Gao2005}, whereas in our case this combination results in repeated bouncing. 

Figure \ref{fig4}b shows that, for all liquids, the diameter $D_2$ of Drop2 decreases linearly with time due to evaporation, until coalescing at $t=0~\rm{s}$ with coalescence size $D_c \sim 1-2~\rm{mm}$. This linear dependence, of the form: $d D(t)/dt=-k$, with slope $k$ dependent on the liquid evaporation rate (gray lines), is in agreement with Ref \citep{maquet2016leidenfrost} for the evaporation of individual ethanol drops in Leidenfrost state, and it contrasts with the power-law model derived in Ref. \citep{biance2003leidenfrost} for single water drops. Furthermore, $D_c$ is largely independent of the initial volume of Drop2 in the range $V_0\in[100,700]\mu l$, as shown in Fig. \ref{fig4}c (solid points). The dashed red line indicates the reference diameter $D_0$ of a spherical drop of initial volume $V_0$; the drops evaporate and decrease linearly in size from $D_0$ until reaching a size $D_c$, of the order of the liquids capillary lengths, $\lambda_c =(\sigma /\rho g)^{1/2}$ (see values in Table \ref{Table1}). The initial volume of the larger central drop (Drop1)  is also not relevant for the coalescence size of Drop2. Figure \ref{fig4}d shows the case of methanol droplets of different initial volumes coalescing with water drops of  0.5, 1.0 and 1.5 ml. The coalescence time for a given $V_0$  is the same in the three cases, indicating that $t_C$ only depends on the volume of Drop2, i.e., when it reaches a given diameter $D_2=D_c$ regardless of the diameter of the central drop $D_1$ (the equivalent coalescence diameter is  $D_{eq}=D_1 D_2/(D_1+D_2)$, since $D_2 \ll D_1$, then $D_{eq} \sim D_2$, i.e., the smaller drop essentially interacts with a flat surface).  Remarkably, $D_c$ is also independent of the substrate temperature (for $T_s >T_L$) as it is shown in Fig. \ref{fig4}e for experiments  performed in the range of $220^{\circ}$C$< T_s <450^{\circ}$C. Although only the case of methanol-water is shown in Figs. \ref{fig4}d-e, similar results were obtained using other liquids. 

\begin{figure}[ht!]
\includegraphics[width=8.5cm]{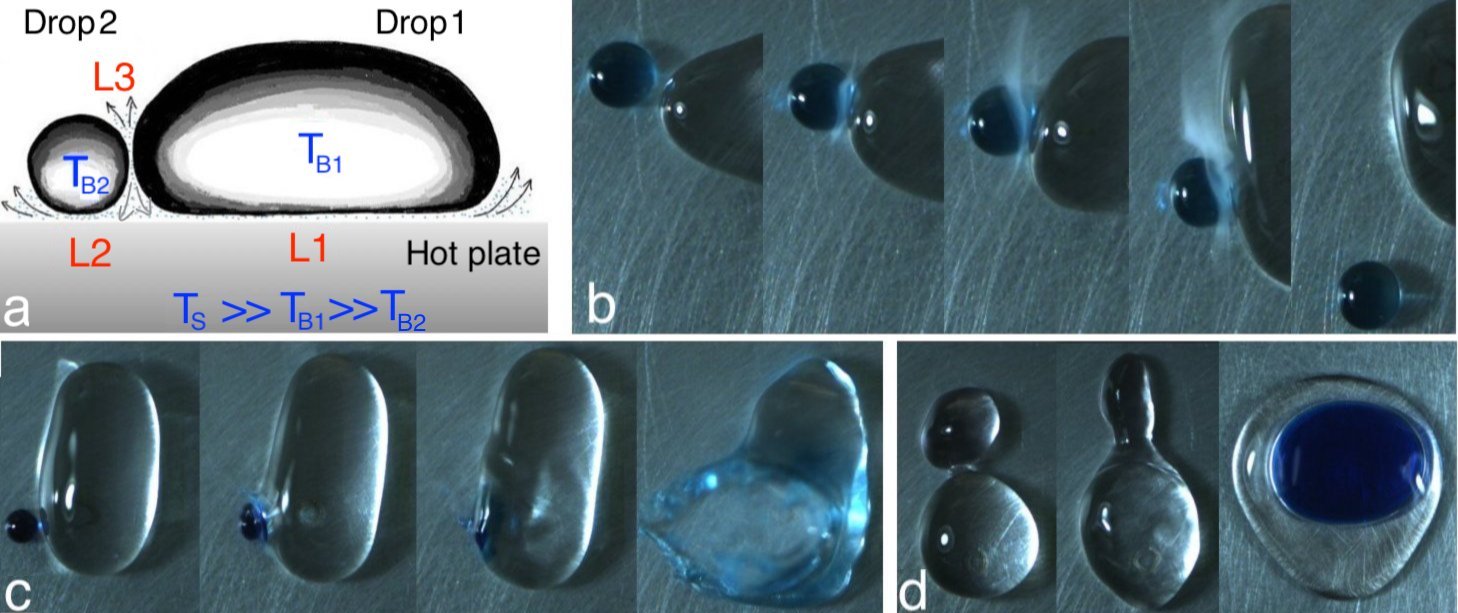}
\caption{a) Sketch of the triple Leidenfrost effect (see text). b) Snapshots taken at 500 fps showing the vapor layer generated when droplets of ethylene glycol (transparent) and chloroform (blue) collide ($T_s=350^{\circ}$C).  c) Explosive coalescence of chloroform (blue) with ethylene glycol. d) Together but not scrambled: water (blue) and toluene join but remain unmixed. }
\label{fig5}
\end{figure}

\paragraph*{The triple Leidenfrost effect.-} Let us now put the pieces together. For two drops to coalesce they must come into contact. If the collision time is very short, the thin gas layer between the droplets is not totally drained and prevents their coalescence. This description fails in explaining why two Leidenfrost drops of the same liquid coalesce directly whereas drops of different miscible liquids do not. We propose an alternative mechanism illustrated in Fig. \ref{fig5}a:  when two droplets are deposited on a very hot plate at temperature $T_s$, both droplets levitate on their own vapor, experiencing independently the well known Leidenfrost effect. The temperature of each droplet is practically its boiling temperature \cite{biance2003leidenfrost};  in the sketch, Drop1 is at temperature $T_{B1}$, and Drop2 at temperature $T_{B2}$. For droplets of different liquids, one droplet has a larger boiling point than the other one. Let say $T_{B1} \gg T_{B2}$. As in the case of droplet levitation on a liquid pool \citep{maquet2016leidenfrost}, Drop1 acts as a superheated surface for Drop2, and consequently, the Leidenfrost state is also established between the droplets, preventing coalescence. Therefore, there are three simultaneous zones of vapor-mediated levitation, indicated by L1, L2 and L3 in Fig. \ref{fig5}a.  This also explains why two droplets of the same liquid coalesce directly, because in such a case $T_{B1}=T_{B2}$ and the Leidenfrost effect between the droplets is not established. Moreover, if $T_{B1} \sim T_{B2}$, the amount of vapor produced during the collision is not enough to avoid contact  \cite{Note3}. The vapor expelled from the bottom of the droplets cannot be the responsible of the frustrated coalescence, because if that were the case, two droplets of the same liquid would also bounce. Finally, when the smaller droplet reaches a size similar to its capillary length, it becomes spherical and the vapor layer in the L3 zone can be easily evacuated, allowing coalescence

Visualizing the vapor layer is a difficult task. Thus, we decided to perform an experiment using liquids with the largest difference in boiling temperatures, and using a Drop2 of the most volatile liquid (lowest latent heat) to produce visible amounts of vapor during the collision.  Based on Supp. Table I \cite{SI}, we used ethylene glycol ($T_{B1}=190 ^{\circ}$C) and chloroform ($T_{B2}=54 ^{\circ}$C,  $L=247$ kJ/kg). Figure \ref{fig5}b shows snapshots of the collision (see also Movie \cite{SI}).  Clearly, a vapor layer is produced between the drops at the moment of collision, preventing coalescence. We presume that the same mechanism is behind the frustrated coalescence for other pairs of liquids, but the amount of vapor produced is scarce and cannot be easily visualized.  When the chloroform droplet decreases in size after several bounces and becomes spherical (i.e. $D_2\sim\lambda_c$), its vapor can easily escape, and coalescence finally occurs. When this happens, $\Delta T$ is so large that the chloroform explodes violently (see Fig. \ref{fig5}c).

Finally, let us describe the particular behavior of toluene and water indicated by ``s" in Table \ref{Table1}.  These liquids are immiscible \cite{Merck2020}, but  they ``merge", with toluene covering the water drop, see Fig. \ref{fig5}d.  Methylene blue in water allows to confirm that there is no mass transfer between the droplets. This contrasts with the case of miscible drops, where the transparent droplet becomes bluish only some milliseconds after the coalescence (Fig. \ref{fig1}d). Perhaps, an inverted Leidenfrost state is established in this case \cite{Vakarelski2012}. This scenario is left for further research.

Summarizing: Two Leidenfrost drops of different liquids bounce one against the other when the difference in boiling points is large, because enough vapor can be generated between them during the collision time. The vapour layer is easily drained when one droplet becomes spherical, allowing coalescence. This mechanism has potential applications as a selective method for droplet separation and coalescence, as we show in Supp. Movie \cite{SI}.

\paragraph*{Acknowledgements.- }Research supported by Conacyt Mexico through Frontier Science 2019 Program (Project: 140604), and C\'atedra Marcos Moshinsky 2020.

\end{document}